\documentclass[10pt, conference, compsocconf]{IEEEtran}
\usepackage{graphicx}
\usepackage{comment}
\usepackage{soul}
\usepackage{amsmath}
\usepackage[textsize=scriptsize, textwidth=7em]{todonotes}
\graphicspath{ {./images/} }

\begin{document}

\title{Predicting Malicious Insider Threat Scenarios Using Organizational Data \\and a Heterogeneous Stack-Classifier}

\author{
\IEEEauthorblockN{Adam James Hall}
\IEEEauthorblockA{Blockpass Identity Lab\\
Edinburgh Napier University\\
Edinburgh, UK\\
adam.hall@napier.ac.uk}
\vspace*{-1cm}
\and
\IEEEauthorblockN{Nikolaos Pitropakis}
\IEEEauthorblockA{Blockpass Identity Lab\\
Edinburgh Napier University\\
Edinburgh, UK\\
n.pitropakis@napier.ac.uk}
\vspace*{-1cm}
\and
\IEEEauthorblockN{William J Buchanan}
\IEEEauthorblockA{Blockpass Identity Lab\\
Edinburgh Napier University\\
Edinburgh, UK\\
w.Buchanan@napier.ac.uk}
\vspace*{-1cm}
\and
\IEEEauthorblockN{Naghmeh Moradpoor}
\IEEEauthorblockA{Blockpass Identity Lab\\
Edinburgh Napier University\\
Edinburgh, UK\\
n.moradpoor@napier.ac.uk}
\vspace*{-1cm}
}

\maketitle

\begin{abstract}
Insider threats continue to present a major challenge for the information security community. Despite constant research taking place in this area; a substantial gap still exists between the requirements of this community and the solutions that are currently available. This paper uses the CERT dataset r4.2 along with a series of machine learning classifiers to predict the occurrence of a particular malicious insider threat scenario - the uploading sensitive information to wiki leaks before leaving the organization. These algorithms are aggregated into a meta-classifier which has a stronger predictive performance than its constituent models. It also defines a methodology for performing pre-processing on organizational log data into daily user summaries for classification, and is used to train multiple classifiers. Boosting is also applied to optimise classifier accuracy. Overall the models are evaluated through analysis of their associated confusion matrix and Receiver Operating Characteristic (ROC) curve, and the best performing classifiers are aggregated into an ensemble classifier. This meta-classifier has an accuracy of \textbf{96.2\%} with an area under the ROC curve of \textbf{0.988}. 

\end{abstract}


\IEEEpeerreviewmaketitle

\vspace*{-0.2cm}
\section{Introduction}
\vspace*{-0.2cm}
Malicious insider threat (MIT) is defined as someone who is motivated to adversely impact the mission of an organization with respect to the confidentiality, integrity or availability of information using the privileges associated with their role \cite{Pitropakis2017}. Insider attack makes up a considerable portion of the cyber-threat landscape, with around 40\% of organizations labelling the vector as the most damaging faced \cite{SansIntitute2017}, and that malicious insiders and hackers make up 47\% of data breaches \cite{Ponemon2017}. MIT is also the most costly per record to resolve (\$155.6 per leaked record \cite{Ponemon2017}). 


Despite the frequent occurrence of IT attacks, detection and mitigation remain problematic. In 2018, 90\% of companies considered themselves vulnerable to insider threats \cite{Associates2018}. A further 38\% of companies admit that their detection and prevention capabilities are inadequate \cite{SansIntitute2017}, and which demonstrates a substantial gap between the current advancements in MIT detection and  growing security requirements. Given the availability of computational resources, using Machine Learning (ML) techniques to solve problems of larger complexity is increasingly a viable option.


As a field, data-driven approaches to detecting MIT is increasing, but front-line attempts still report more effective models than where machine learning has been applied \cite{Gheyas2016}. These initial attempts have often used a Graph-Based approach \cite{Eberle2009} and fuzzy logic-based anomaly detection approaches \cite{Yingbing2007}.

This paper presents a new methodology for processing organizational log data into a format for classifying whether particular individuals belong to a particular threat archetype on a daily basis. It then outlines the training of multiple learning algorithms in order to classify this threat scenario, while experimenting with boosted and non-boosted learning methods. The best performing algorithms are aggregated using a probability vote in order to create a model which has the largest area under ROC curve of all the developed models.

The contributions of our work can be summarized as:
\begin{itemize}
  \item A methodology for splitting MIT into subcategories to improve predictive performance when compared to previous prediction approaches, which largely treat insider threat as a single category. MIT can take a number of forms, this can complicate prediction for these techniques. The present work identifies a model by which individual threat archetypes are detected through supervised learning algorithms.
  \item Investigation of boosting when optimizing the performance of classification algorithms in the field of application.
  \item Demonstrates an approach for aggregating high-performance classifiers into an optimal meta-classifier in the MIT domain.
\end{itemize}

The rest of the paper is organized as follows: Section II provides a related literature review, Section III introduces the MIT scenario addressed by the approach. Section IV presents our approach to detecting malicious activities in synthetic organizational records. Section V offers an evaluation of our proposed model and, finally, Section VI draws conclusions giving some pointers for future work.

\section{Related Work}
\vspace*{-0.2cm}

The most popular approach for dealing with MITs remains unsupervised, anomaly-detection based approaches \cite{8411913}. One unsupervised approach - and which was applied to the CERT dataset r6.2 - uses a deep neural network to establish a baseline of normal behaviour for each user for each day and compare to new days against. When anomaly scores are organized into employee percentiles for each day, almost every malicious employee is placed well above the 95th percentile for high anomaly scores \cite{Tuor2017}. This shows great potential in having the ability to quickly create a shortlist of high-risk individuals. However, these systems often lack the capacity to identify positive or negative cases of insider threat. They also give no indication as to the nature of the anomaly, whether it could be a malicious event or it could just be caused by employees breaking their usual work habits in innocuous ways. The task then falls on the security operations center (SOC) to classify the nature of the anomaly. This, though, can be laborious and time consuming.

These drawbacks can be mitigated when using classification techniques to identify insider threat by training specific models to identify particular attack scenarios. Despite this potential of breaking MIT into smaller categories for prediction, this approach has been rarely identified in the literature. Instances of classifiers being used in research to predict MIT tend to treat MIT as a single class of problem.  In addition, data sources used for MIT classification can vary. Most of the data sets used in related classification problems use log information from the systems that individuals have accessed. This can be further sub-categorised into synthetic, non-synthetic and mixed log data. However, recent research also shows data being analyzed from a plethora of different sources.

One example of an approach that breaks the convention of predicting MIT from logs uses psycho-physiological signal data to train Support Vector Machines (SVMs). This is taken from electroencephalography and electrocardiogram sensors which are placed on a small group of participating individuals who either performed an intentional MIT activity or were benign. This data is then used to train SVMs to classify instances with an average accuracy of 86\% \cite{Hashem2016}. While this approach was able to perform with a reasonable accuracy, it is difficult to say whether this would be true in a non-staged MIT environment. In addition, the sample size was only 10, and this would have to be tested on a much larger population sub-sample in order for the resultant classifier to be credible. Finally, even if the equipment were available to feasibly perform this kind of analysis, there may be further obstacles when acquiring consent from employees around undergoing full-time analysis at work.

Another example that appears in the research uses both classification and clustering techniques on real-world organizational data \cite{Gavai2015}. This two-pronged approach attempts to predict which employees in the organization are likely to quit using classification while also using an unsupervised approach to detect which users may be insider threats. The classifier had an accuracy of 73.4\% when detecting quitters. The unsupervised approach for detecting insider threats was effective in that all insider threat cases had an anomaly measure of above the median score. However, this tended to be the norm among the scores of benign individuals also, casting doubt on the effectiveness of this approach for predicting insider threat as a single class.

One final, single-pronged approach, creates a classifier using the CERT synthetic dataset r6.2 to predict insider threat. Here researchers compare the performance of traditional machine learning algorithms against their long short-term memory recurrent neural network. This classifier achieved an accuracy of 93.85\%, outperforming the next most accurate algorithm by around 5\% \cite{8411913}. This accuracy was achieved through thoroughly pre-possessing the initial log data. Firstly, events are standardized and aggregated into a format around the behaviours and attributes of individuals. Features are then extracted for the training phase and testing phases respectively.

Our methodology takes into consideration the other approaches of MIT detection proposed in the literature but it expands upon this through introducing boosting, stacked classifiers and the use of behavioural archetypes to narrow down on the scope of prediction.
\vspace*{-0.23cm}
\section{Malicious Insider Threat Model}
\vspace*{-0.2cm}
Following the prevalence of previous approaches in predicting insider threats in the CERT synthetic dataset ecosystem \cite{8411913} \cite{Tuor2017}, this source was chosen to perform our experiments. However, instead of choosing r6.2 with only one instance of each threat, r4.2 was chosen. Unlike r6.2, r4.2 was created as a \emph{dense needle} data set. This contains a large number of positive cases of each threat scenario. This is an ideal classification problem as there is a vast wealth of positive cases which predictive models can learn the structure of. With more true cases to learn from, the resultant classifier is likely to be far more adept at correctly identifying true future cases.

Dataset Version 4.2 contains around 20GB of employee activity logs for 1000 employees over 17 months. Three different attack scenarios have been simulated, each allocated 30 malicious employees from the 1000 employee pool. These are described as follows:

\begin{enumerate}
  \item User who did not previously use removable drives or work after hours starts to logging in after hours, using a removable drive, and uploading data to \emph{wikileaks.org}. Leaves the organization shortly thereafter.
  \item User begins surfing job websites and soliciting employment from a competitor. Before leaving the company, they use a thumb drive (at markedly higher rates than their previous activity) to steal data.
  \item System administrator becomes disgruntled. He downloads a key-logger and uses a thumb drive to transfer it to his supervisor's machine. The next day, he uses the collected key logs to log in as his supervisor and send out an alarming mass email, causing panic in the organization. He leaves the organization immediately.
\end{enumerate}

Despite previous research \cite{8411913}, the authors in this work did not focus on creating a general-purpose model for predicting MIT regardless of the MIT scenario category. As the three scenarios provided suggest, MIT can take many forms. Not all scenarios will carry a similar signature. Any model which is applied to MIT as a blanket solution will need to be vague enough not to rule out MIT cases which are distinct in nature, but also well-fitted enough to actually catch cases of MIT in their particular scenarios. One model per scenario ensures that even small nuances in each case are learned, whereas a more generalized model may become less accurate if classifying threats across a wider spectrum. If this is successful, the dependent variable can be turned into a categorical variable to predict each distinct scenario. However, this categorical approach may not be compatible with some kinds of learning algorithms and may also create unnecessary noise for algorithms to deal with. A potentially better approach would be to train a separate model to detect each separate threat scenario. This would allow each model to be as well fitted to its specific scenario as possible.

Data set r4.2 is originally composed by the following sets of logs: 1) Employee Login/Logoff event logs; 2) Device Connect/ Disconnect event logs; 3) Employee HTTP event logs; 4) Monthly Record of Employees; 5) Psychometric Profiles of Employees; and 6) File Accesses.
\vspace{-0.25cm}

\section{Methodology}
\vspace*{-0.1cm}
\begin{figure*}[!htb]
\centering
\fbox{\includegraphics[width=6in]{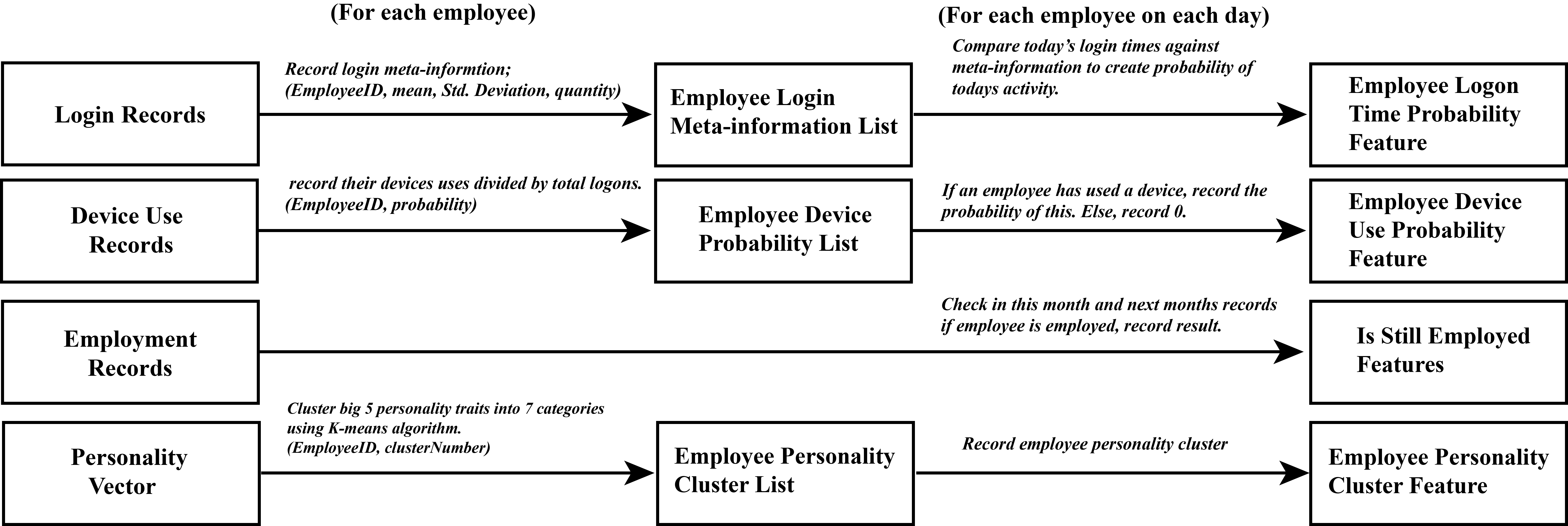}}
\DeclareGraphicsExtensions.
\caption{Architecture of data pre-processor}
\label{fig_sim}
\vspace{-0.35cm}
\end{figure*}

\subsection{Data Aggregation and Feature Selection}

In order to decide which elements should be aggregated to generate features, threat scenario one is further analyzed. Three key traits of the scenario are identified: 1) Users who are not usually device users suddenly start using a device; 2)Users who usually only log in during work hours start logging on after work hours; and 3)Users leave the organization shortly after the incident.


Data is aggregated from the logs to give a summary of each individual in the organization with respect to the information identified. This is done in daily time intervals as this provides a good balance between time-frame granularity and computational complexity of training. The approach is outlined in Figure 1.  With this approach, the aggregated training data will take the form of a daily summary of each employees activity. As only employees who have used a device are capable of performing this MIT scenario, only these employees have been included in the training data. This reduces the number of employees from 1000 to 266. In addition, in order to reduce training time, only data from the month of July has been added to the set. This month was chosen because it had the highest incidence of MIT. After aggregating data for each employee who was active during the month of July, we have a base training set of 7260 instances where only 18 are positive threat cases. In order to reduce this set imbalance, a spread sub-sample of negative cases is taken. This reduces the negative to positive ratio to only 15 to 1, leaving a training set size of only 288 instances. The elements of this training data are the features selected during the pre-processing phase.

The first trait of scenario one is an abnormality of device usage. The probability of a device being used is calculated for each employee using the information in the device connection logs. Each employees probability of using a device ($P(D)$) is derived using the Formula in Equation 1. $U$ is the number of device connects associated with a user and $T$ is the total number of connect events in the log. This probability for each employee is stored into a list. If an employee uses a device on any day, the probability of this happening is recorded as a feature. If there is no device usage, this is recorded as 0.
\vspace{-0.25cm}

\begin{equation}
P(D) = \frac{U}{T} \qquad Z = \frac{\bar{X} - \mu}{\left ( \frac{\sigma}{\sqrt{n}} \right )}
\end{equation}

The distribution of threat cases with regard to the probability of an employee using a device is shown in Figure 2.
\vspace{-0.45cm}
\begin{figure}[!htb]
\centering
\fbox{\includegraphics[width=3.2in]{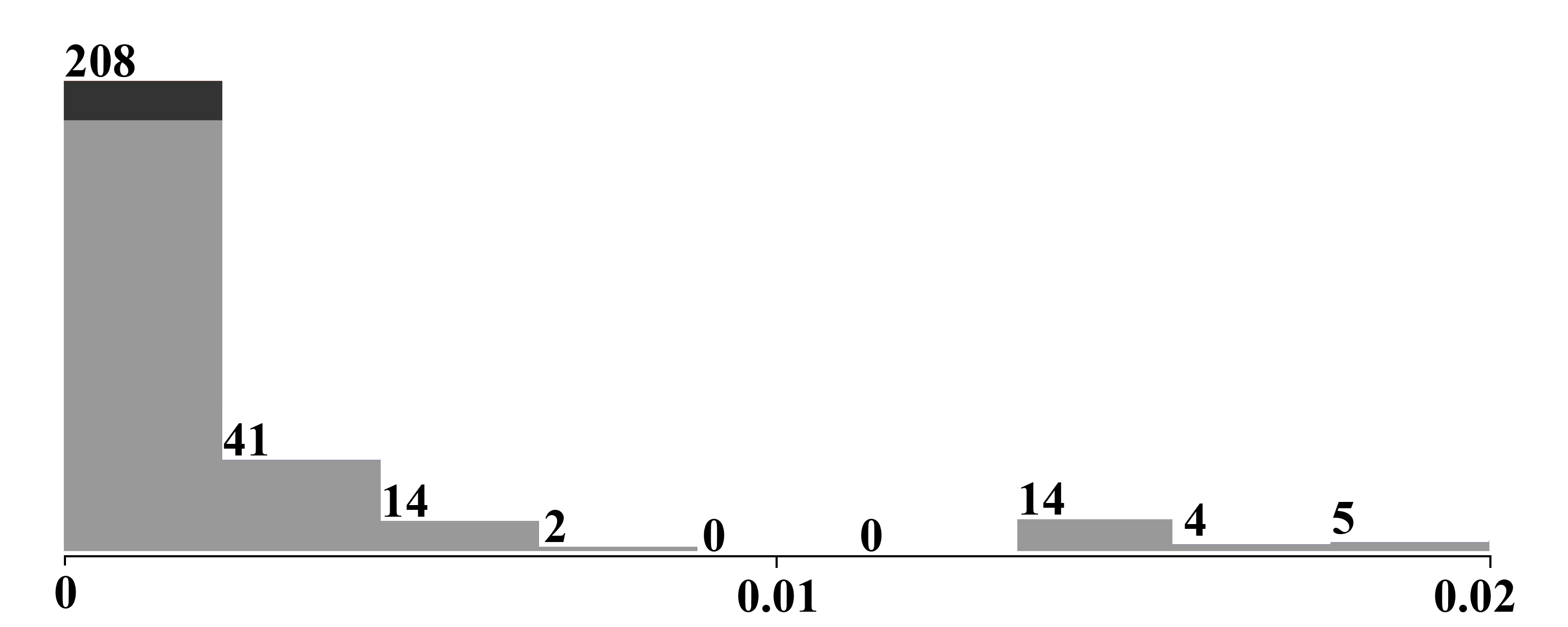}}
\DeclareGraphicsExtensions.
\caption{Histogram showing distribution of threat cases (shown in dark grey) among the probabilities of employees using a device}
\label{fig_sim}
\vspace{-0.35cm}
\end{figure}

The second signature of scenario one is an abnormality in logon times. In order to turn this into a feature, a probability distribution is generated for the log in times of each employee. This is achieved for each employee through the following steps; All of the logon times for each individual employee is compiled into a list. The mean (\( \mu \)) and standard deviation (\( \sigma \)) and number of logon time measurements ($n$) are recorded for each employee. These can then be used with each new logon time on each new day (\(\bar{X}\)) for each employee to create Z-scores (Z). This equation is shown in Equation 2. These Z-scores can be plotted onto a normal distribution curve using a Z-table. 


This gives us the probability that any employee will log in at any particular time, with respect to their personal habits. The distribution of threat cases with regard to the probability of an employee logging on at that time is shown in Figure 3. 

\begin{figure}[!htb]
\centering
\fbox{\includegraphics[width=3in]{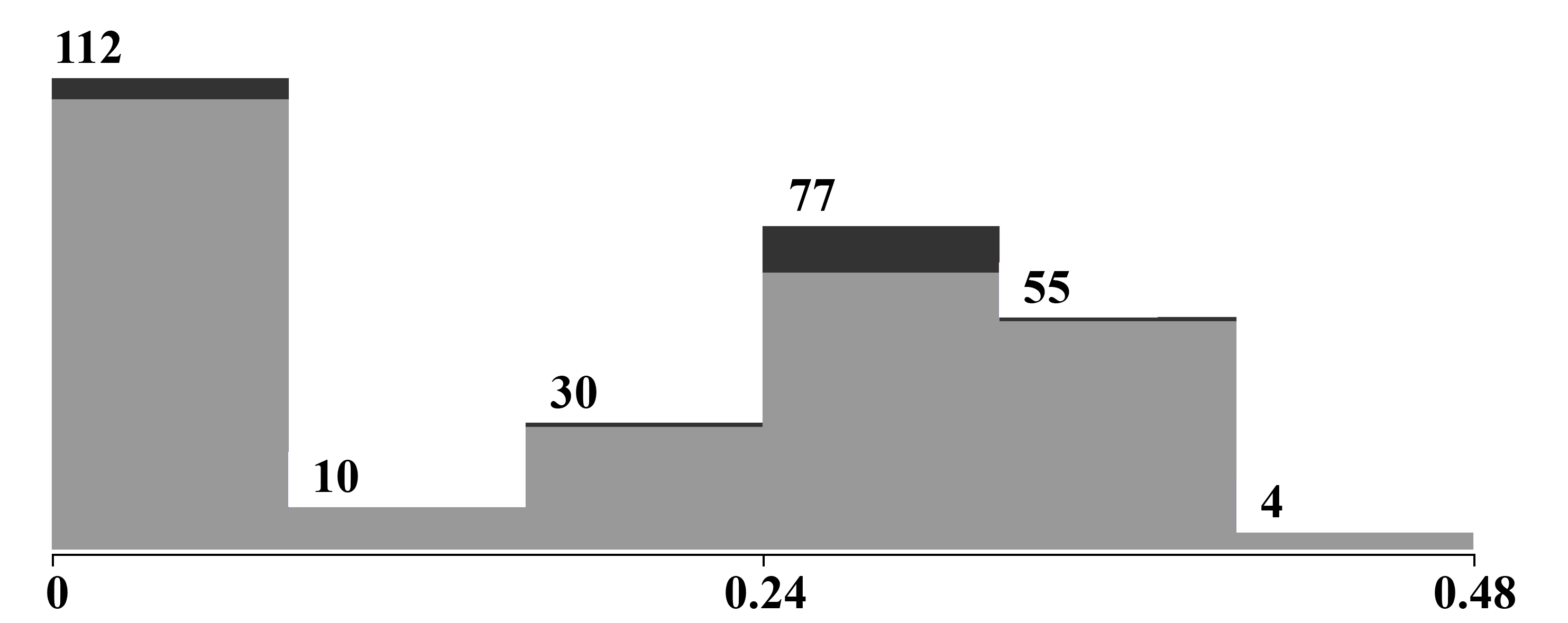}}
\DeclareGraphicsExtensions.
\caption{Histogram showing distribution of threat cases (shown in dark grey) among probability of an employee logging on at a particular time}
\label{fig_sim}
\vspace{-0.75cm}
\end{figure}

The third identified trait of this threat scenario is employees leaving the organization shortly after the incident. Employee records are supplied in our data set. If the employee isn't in the records for any particular month, this means that the employee is no longer employed in the organization. In order to add this as a feature to our training data, when a new day instance is being aggregated, the employee logs for the current month and the next month are searched for that user ID. The positive or negative results of this search are then recorded as features for that user on that day. In the data, all of the positive cases for this scenario were not employed next month.


While the psychometric information was not directly included in the threat scenario traits, this is still added as a feature in the training data in order to test for significance. Originally, this takes the form of a vector denoting the employees score on the 'Big 5 Personality traits' indexes. These vectors were clustered into seven categories using a simple k-means algorithm. The personality cluster for each employee was then recorded into a list to be referenced when generating the training data. This can be observed in Figure 4, true MIT cases only appear in four of the seven psychometric categories. 

\vspace{-0.25cm}
\begin{figure}[!th]
\centering
\fbox{\includegraphics[width=3.2in]{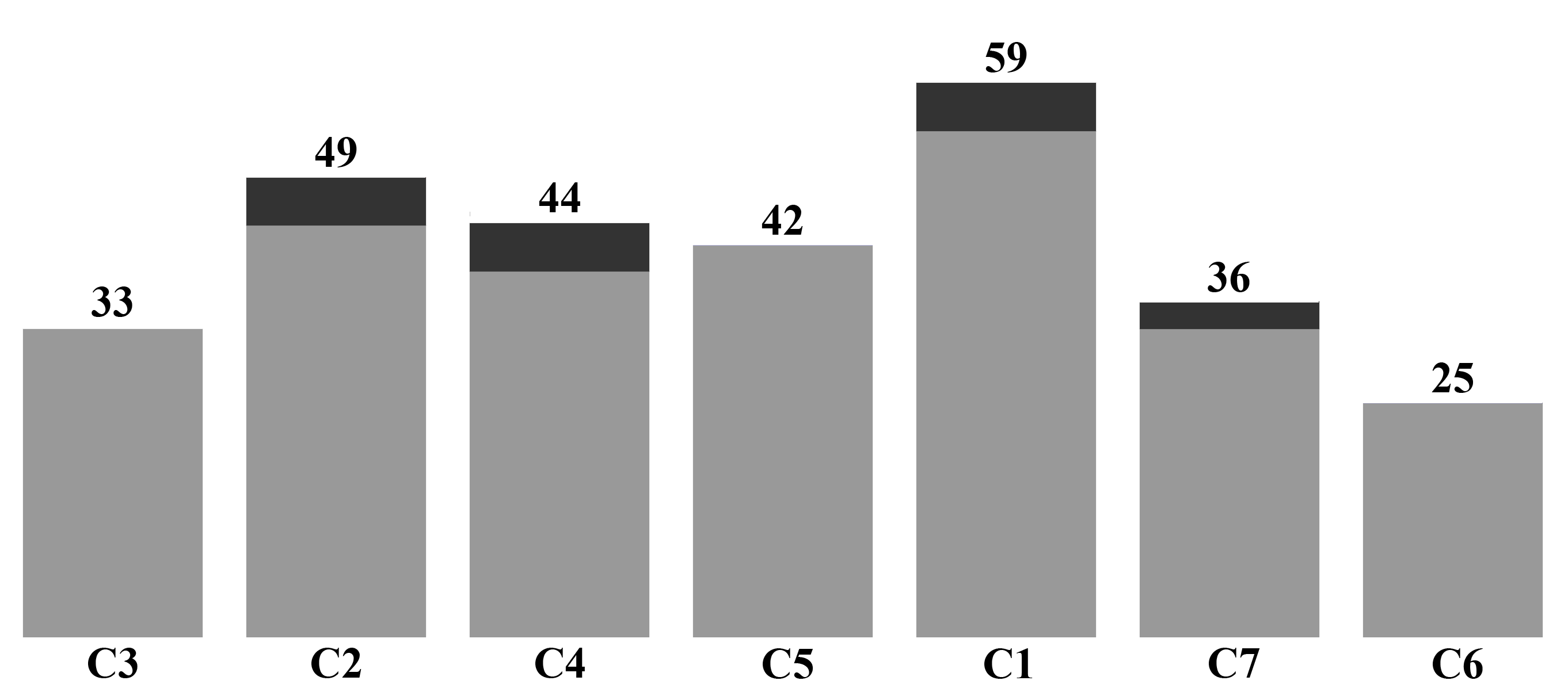}}
\DeclareGraphicsExtensions.
\caption{Histogram showing distribution of threat cases (shown in dark grey) among psychometric clusters}
\label{fig_sim}
\vspace{-0.35cm}
\end{figure}
\vspace{0.2cm}
Finally, labeled threat cases are taken from the answers section of data set version 4.2.  The format of each instance in the set is as follows: I)Employee log on time probability \textit{(continuous)}; II)Device Connect probability \textit{(continuous)}; III)Is the employee employed this month? \textit{(boolean)}; IV)Is the employee employed next month? \textit{(boolean)}; and V)Psychometric cluster of employee \textit{(categorical)}.


\subsection{Model Building}

In order to build models on this training data, the Weka toolkit was utilized \cite{Weka}. Using this toolkit, multiple learning algorithms were trained on the data. Each algorithm was then compared against a boosted version of the same algorithm. Boosting is where classifiers are trained iteratively, at each iteration the incorrectly classified instances are amplified in the training data in order to, ideally, improve performance. After boosting, the resultant accuracy's and the area under the Receiver Operating Characteristic (ROC) curve were compared to evaluate model performances. These included: 1) Neural Network (NN); 2) Naive Bayesian Network (NBN); 3) Support Vector Machine (SVM); 4) Random Forest (RF); 5) Decision Tree (DT); and 6) Logistic Regression (LR).


Results are shown in Tables I and II. Models are validated using 10-fold cross validation \cite{Wooldridge2018}
\vspace{-0.25cm}
\begin{table}[ht]
\caption{Classifier Accuracy's before boosting} 
\centering 
\begin{tabular}{c c c c c c c} 
\hline 
Classifier & NN & NBN & SVM & RF & DT & LR \\ [1ex] 
\hline 
Accuracy & 95.8\% & 91.3\% & 95.4\% & 97.5\% & 96.1\% & 96.5\%\\ 
Area under ROC & 0.974 & 0.954 & 0.872 & 0.982 & 0.915 & 0.983 \\
 [1ex] 
\hline 
\end{tabular}
\label{table:nonlin} 
\vspace{-0.25cm}
\end{table}

\begin{table}[ht]
\caption{Classifier Accuracies after boosting} 
\centering 
\begin{tabular}{c c c c c c c} 
\hline 
Classifier & NN & NBN & SVM & RF & DT & LR \\ [1ex] 
\hline 
Accuracy & 95.8\% & 97.2\% & 97.2\% & 97.2\% & 97.2\% & 96.8\%\\ 
Area under ROC & 0.952 & 0.980 & 0.980 & 0.888 & 0.932 & 0.802 \\
 [1ex] 
\hline 
\end{tabular}
\label{table:nonlin} 
\vspace{-0.55cm}
\end{table}

The results of boosting are mixed. Some are improved by the boosting approach and some do not perform as well. The best performing classifiers are selected to be aggregated into a heterogeneous stack classifier. The best performers are identified using the ROC and accuracy values shown in Tables I and II. These algorithms are: 1)Neural Network; 2)Boosted naive Bayesian Network; 3)Boosted Support Vector Machine; 4)Random Forest; and Logistic Regression

The aforementioned algorithms are aggregated into a single \emph{metalearner} using probability vote. By combining algorithms in this way, we can leverage the strengths of separate models to approximate a greater area under the ROC \cite{Witten2013}. In Figure 5, the shaded area outside of both of the functions is approximated by combining methods A and B. Similarly, the area between the ROCs of the five identified models is approximated. This is achieved by combining the classifiers using probability vote, allocating the vote weight based on the probability of each classifier being correct. This learner missed only correctly classified 14 out of 18 true cases and 262 out of 270 false cases. 

\vspace{-0.15cm}
\begin{figure}[!th]
\centering
\includegraphics[width=3.2in,scale=0.3]{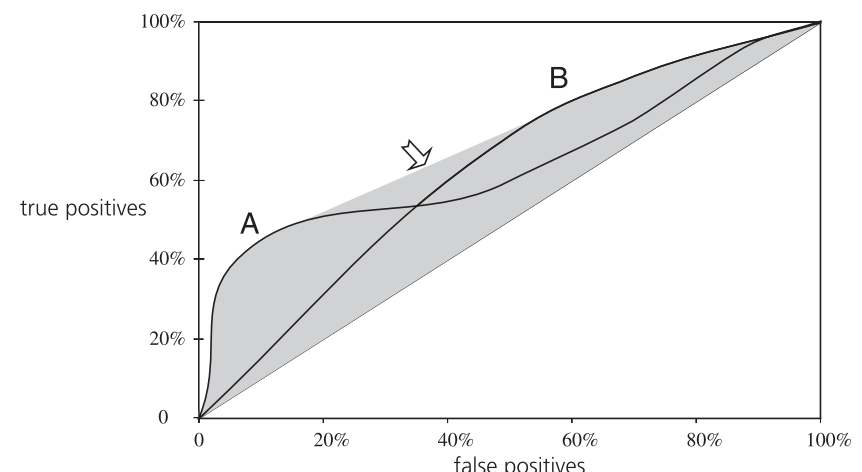}
\DeclareGraphicsExtensions.
\caption{Graph  demonstrating performance approximation of models A and B \cite{Witten2013}.}
\label{fig_swim}
\vspace{-0.45cm}
\end{figure}

In Table III, the performance of the meta-learner is compared against the next best performing classifier; a boosted naive Bayesian network. In Figure 7, we see the ROC curve of the meta-learner.


\vspace{-0.25cm}
\begin{table}[!th]
\caption{Metalearner Performance Comparison} 
\centering 
\begin{tabular}{c c c c} 
\hline 
 & MetaLearner & Boosted NBN & Difference \\ [1ex] 
\hline 
Accuracy & 96.2\% & 97.2\% & -1\% \\ 
Area Under ROC & 0.988 & 0.980 & 0.08\% \\
 [1ex] 
\hline 
\end{tabular}
\label{table:nonlin} 
\vspace{-0.3cm}
\end{table}

\begin{figure}[!htb]
\centering
\includegraphics[scale=0.25]{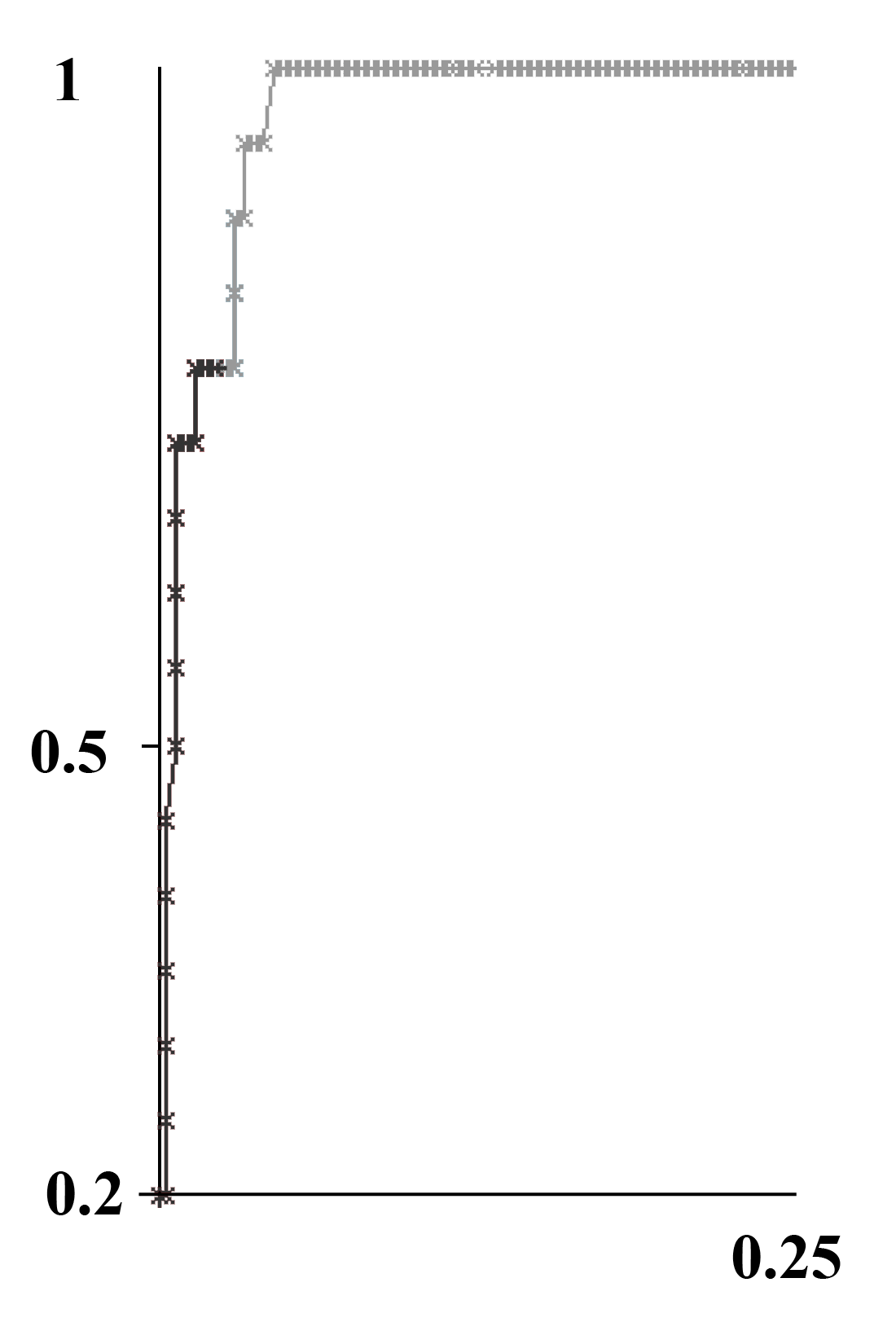}
\DeclareGraphicsExtensions.
\caption{ROC curve showing predictive performance of the meta-learner. MIT cases are in dark grey, non-MIT cases are in light grey.}
\label{fig_sim}
\vspace{-0.65cm}
\end{figure}

\section{Evaluation}

\subsection{Pre-processing}

During the course of this work, one archetypal threat scenario was analyzed and a data pre-processing approach was developed to optimize instance data for predicting this scenario. While the original data set was large and complex, this was significantly distilled in order to create instances out of just five data points. The histograms representing the instance attributes in figures 2 to 5 each show a discernible split between the MIT and non-MIT cases. The quality of these features is demonstrated by the performance metrics of the algorithms that trained upon them. Each classifier that trained upon the data had an accuracy of at least 95\%, which is highly accurate. The only common factor present in training the classifiers that produced these performance metrics was the data created during the pre-processing phase. 

\vspace{-0.22cm}
\subsection{MIT Category Granularity vs. Workload Trade-off}
\vspace{-0.25cm}
Despite the quality of training data, our approach addresses only one threat archetype at the present time. The authors intend to expand this; developing alternative models for separate categories. Using this approach, similar pre-processing will have to be performed for each MIT scenario. This represents a greater pre-processing workload than approaches that treat MIT as a single category. By extension, this is also true for any learning algorithms that will be trained. Separate classifiers will have to be trained for separate archetypes.

In order to combat the increase in workload associated with the finer granularity in MIT category, different levels of archetypal nuance can be established in order to optimize the number of archetypes to the associated pre-processing and training workload. In the training data, there were only three archetypes. This is likely to be different in real-world scenarios. In real data, there is likely to be an entire spectrum of insider threat scenarios. This spectrum, however, could be split into a number of subcategories, where scenarios in each category carry similarities. The more times that MIT has been divided into subcategories.  

The more work that will be involved in the pre-processing as more training data sets will need to be developed and subsequently more models will need to be trained on these separate sets. Having said this, on the basis of the results observed in this study, it can be hypothesized that the greater the nuance in MIT category, the more accurate the learner is likely to be. 

\subsection{Metalearner Performance}

While the accuracy of prediction is not improved through the aggregation of high-performance models into a probability vote meta-learner, the Area under the ROC is the greatest of all that have been observed. This shows that, in respect to this metric, the meta-learner is greater than each of its constituent classifiers. While the overall accuracy suffers in this approach, this is only due to the fact that a small number of negative cases have been classified as true. The large area under the ROC shows that more true cases have been identified correctly than any other model. The cost of incorrectly identifying an insider as MIT is the cost of a SOC analyst verifying that this insider has not, in fact, performed one of the threat scenarios. The cost of incorrectly identifying a true insider case as negative is the cost of the data breach. Depending on sensitivity and quantity of records that are leaked, this could cost millions. For this reason, the authors have identified the meta-learner as the most fit for purpose model developed during this research.

\vspace{-0.25cm}
\section{Conclusion}
\vspace{-0.10cm}
In this work, a new methodology for pre-processing insider threat to optimize classification results based on insider threat categories is established. The resultant data set produced using the established methodology is used to train a series of classifiers which all outperform the predictive performance of previous strategies identified in the research. The most performant of these models are aggregated into a meta-learner algorithm using probability vote. This produces a model with a ROC curve containing a greater area underneath than any of the other models that were explored in this work. This indicates the suitability of this approach for improving overall classifier performance. 

On the basis of results identified, this work could be further expanded by tailoring instance data to the other two scenarios present in data set r4.2. A general model could also be developed on this data in order to test the hypothesis that instance data tailored to each scenario creates more performant classifiers than one generalized classifier. In addition these features could possibly be extended using a genetic algorithm approach, this may produce features of higher quality. Finally, real-life data could be used to train future models relating to red-team simulated scenarios. This would allow the effectiveness of this approach to be tested in the wild, further validating it's applicability to this problem domain.

\vspace{-0.15cm}



\bibliographystyle{IEEEtran}
\bibliography{citations}

\end{document}